\documentclass[12pt,preprint,epsfig]{aastex}
\newcommand{\be}{\begin{equation}}
\newcommand{\ee}{\end{equation}}
\newcommand{\bd}{\begin{displaymath}}
\newcommand{\ed}{\end{displaymath}}

\newcommand{\eliK}{{\rm I\!K}}
\newcommand{\eliM}{{\rm I\!M}}

\slugcomment{Submitted to Astrophysical Journal}
\shorttitle{The extrasolar planets HD 82943 c,b}
\shortauthors{S. Ferraz-Mello, T.A. Michtchenko and C. Beaug\'e }
\begin{document}
\title{The orbits of the extrasolar planets HD 82943 c,b}

\author{S. Ferraz-Mello, T.A. Michtchenko}
\affil{Instituto de Astronomia, Geof\'{\i}sica e Ci\^encias Atmosf\'ericas,
       Universidade de S\~ao Paulo, Rua do Mat\~ao 1226, 05508-900 S\~ao Paulo,
       Brasil}
\and
\author{C. Beaug\'e}
\affil{Observatorio Astron\'omico, Universidad Nacional de C\'ordoba, Laprida
       854, (X5000BGR) C\'ordoba, Argentina}

\begin{abstract}
The published orbits of the planets HD 82943b and HD 82943c correspond to a system bound to a catastrophic event in less than 100,000 years.
Alternative sets of elements and masses, which fit the available observational data and correspond to regular motions, are presented in this paper. 
The planets HD 82943 c,b are in a 2/1 mean-motion resonance and are trapped in the neighborhood of a (0,0)-apsidal corotation.
\end{abstract}
\section{Introduction}


This paper is a study of the orbital elements and masses of the planets HD 82943 c,b on the basis of the available data. 
Almost every paper on the dynamics of the actual extra-solar planetary systems assumes given sets of elements and masses (which includes the mass of the star) as a revealed truth. 
Sometimes, arbitrary variations of the revealed values are introduced in the hope of learning more about the systems stability. 
Generally, these studies are impaired by the scarce information given on the way in which elements and masses were obtained.

\begin{figure}[t!]
\epsscale{.80}
\plotone{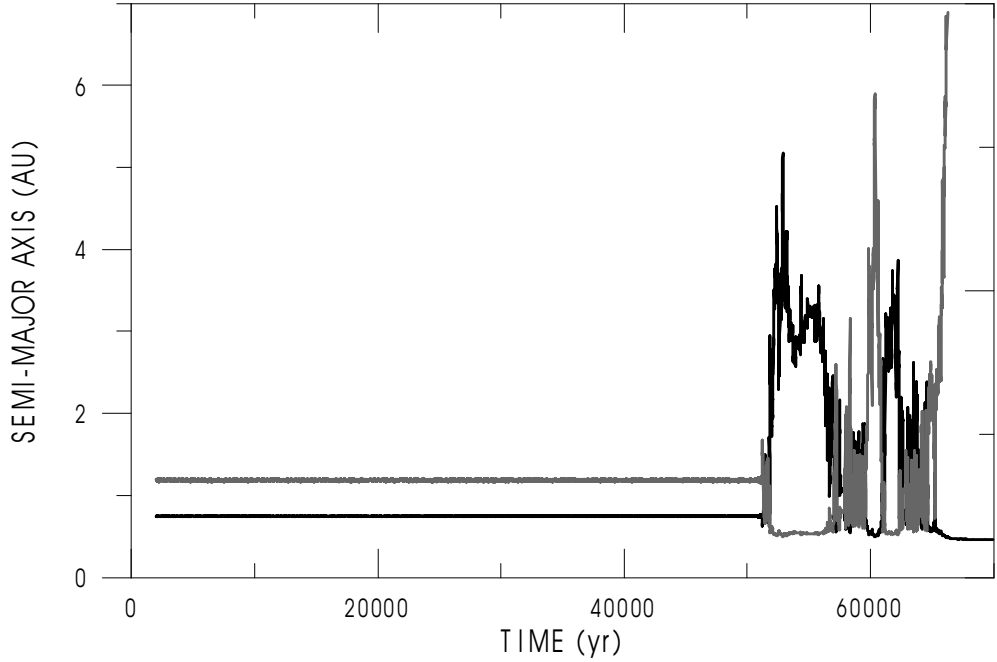}
\caption{Evolution of the semi-major axes of HD 82943 c,b considering the masses ($M_{\star}=1.15 M_\odot, m_1=1.85 M_\mathrm{Jup}, m_2=1.84 M_\mathrm{Jup}$), the orbital elements given by \citet{May} and coplanar orbits.}
\label{catastro}
\end{figure}

The recently published orbits of the planets HD 82943 c,b \citep{May} would have a similar fate would they not correspond to a highly unstable system.
The published orbits undergo a major catastrophe in about 50,000 years and one of the planets is expelled from the system (or collides with the star) in less than twice that time (see fig. \ref{catastro}). 
In the first 10,000 yr of the simulation, the solution lies in the 2:1 resonance zone and one of the two critical angles related to this resonance appears in libration; however, the permanence of the system in a resonant state is only temporary. 
 
For the sake of comparison with similar simulations, we shall emphasize that the result shown in fig. \ref{catastro} is very sensitive to parameters not fixed by the given data as the epoch in which they are valid (which are arbitrary in a Keplerian orbit determination) or the nature of the elements (mean or osculating). It is also sensitive to the actual stellar and planetary masses. 
As it will be discussed later, these masses are only poorly determined. 
The observations allow a function of the stellar and planetary masses and the inclination of the orbital plane to be determined, but this function cannot be solved to give separately each of these parameters.

Anyway, in all simulations done, the fate was similar with, at most, a tenfold variation in the time of the catastrophic events. Some solutions developed instability in less than 5,000 yr! However, HD 82943 is a 2.9 billion-yr old star \citep{May} and we should expect the planets to be stable for timespans of at least that order. Therefore, no instability should manifest in a much shorter time scale, be it a backward or a forward simulation. 
         
Simulations of the primordial disk-planet evolution of pairs of planets in close orbits indicate that a capture into resonance followed by a capture into an apsidal corotation is likely to occur as a consequence of planetary migration \citep[see][]{Pap}. The past evolution should be reflected in the orbital characteristics of planetary systems with a period ratio not larger than 3 or 4, where the gravitational interaction between the bodies is significant. However, the number of systems that could serve as examples of this phenomenon is very small. Only two of the possible resonant systems of exoplanets are confirmed: GJ 876 and HD 82943. Other candidates, as 47 Uma and 55 Cnc are still affected by doubts concerning the very existence of one of the planets \citep{Nae}. The small number of such systems justifies the need of improving our knowledge on the orbits of the planets HD 82943 c,b, notwithstanding the several difficulties discussed thereafter, instead of just waiting for new elements. A better knowledge of the orbits of these planets is necessary to constrain dynamical studies of the evolution of extra-solar planets. 

The results reported in this paper were obtained with a observation series based on data published in graphic form. This is a delicate operation and special care was taken with points located in the steep branches of the radial velocity curve, since a small error in the t-coordinate generates a large $O-C$ ($O=$observed radial velocity; C = radial velocity calculated from the elements) and impairs the results. The radial velocities thus obtained are accurate, but the times of observation are not. It is not possible to get a precision better than 1 day from the plot. Anyway, the r.m.s. of the residuals of the series used in this paper, with respect to a best-fit solution is the same as that of the original series \citep[6.8 m/s cf.][]{May}.

\section{The published orbits}
Let us start this paper discussing the orbits published by \citet{May}. We do not need to emphasize again that they correspond to a highly unstable pair of planets and are likely untrue. We have to emphasize, however, that the given orbits fit the observations very well and that the problem is not that of improving the orbits published by Mayor et al., but of finding alternatives to them. As we are interested in the real planets, we rule out the possibility of introducing any blind variation of the elements just to get stable orbits. The domain of stable resonant orbits is vast and just pointing out one or more ``ad-hoc" stable orbits adds nothing to our knowledge of these planets if these orbits do not fit the existing observations.

The orbits given by Mayor et al. come from the fitting of a pair of Keplerian ellipses to the observations. 
In the dual Keplerian model, the mutual attraction of the planets is disregarded and the barycentric velocity of the star projected on the line of sight is given by
\be
V_R = \sum_k \eliK_k [\cos(f_k+\omega_k) + e_k\cos\omega_k] + V_r
\label{veloc2}
\ee
where 
\be
\eliK_k =\frac{m_k}{\eliM}\,\frac{2\pi \bar{a}_k}{ \widehat{T}_k} \,\frac{\sin i_k}{\sqrt{1-e_k^2}}
\label{eliK}.
\ee
The parameters involved in these equations are the system velocity ($V_r$), 
the star mass ($M$),
the total system mass ($\eliM$) and, for each planet,
the true anomaly ($f$), 
the eccentricity ($e$), 
the distance from the tangent plane to the celestial sphere (sky plane) to the periastron ($\omega$),
the inclination of the orbital plane over the sky plane ($i$), 
the sidereal period ($\widehat T$),
the mean distance to the star ($\bar a$), 
and the planet mass ($m$).
In the published elements, the periastron is measured from the point where the planet orbit crosses the plane tangent to the celestial sphere when moving towards the observer (it corresponds to the ascending node of the barycentric orbit of the star -- near the radial velocity maximum).  

Usually, Kepler's third law is used to eliminate $\bar{a}_k$. 
In the case of just one planet, $\widehat T$ and $\bar a$ are equivalent to their osculating counterparts $T,a$ and we have 
${4\pi^2}{\bar a}^3={G(M+m)}{\widehat T}^2$. 
Using this law to separate, in eqn. (\ref{eliK}), the known and unknown parts, we obtain,
\be
\frac{m^3}{{\eliM}^2}\sin^3 i = \frac{\eliK^3 T}{2\pi G} (1-e^2)^{3/2}. 
\label{1planeta}\ee
where $\eliM = M + m$.
The left-hand side:
\be
f(M,m,i)=\frac{m^3}{{\eliM}^2}\sin^3 i = \left(\frac{m}{\eliM}\sin i\right)^3 {\eliM} 
\ee
is the so-called {\it mass-function}.  \citep[See, for instance,][]{Plu}.
 
When the system has two planets, $\bar{a}_k$ and $\widehat{T}_k$ are no longer the osculating orbital elements and a correction to Kepler's law is necessary. 
We have to adopt a different mass-function:
\be
		f(M,m_k,i_k)=\frac{m_k^3}{{\eliM}^3}\sin^3 i_k (M+m_k).
\ee
We have, also, to distinguish between the total mass of the system, $\eliM$, and the sum of masses $M+m_k$ and it is no longer possible to cancel the last term in the above equation with one power of $\eliM$ as in the case of only one planet.
Instead of eqn. (\ref{1planeta}), we have,
\be
	f(M,m_k,i_k)(1+\frac{1}{2}\sigma_k) =
\frac{\eliK_k^3 \widehat{T}_k}{2\pi G} (1-e_k^2)^{3/2}.
\ee
where $\sigma_k$ are the classical first-order corrections necessary to transform mean to osculating elements (see Appendix). In the case of two planets near the 2/1 resonance, neglecting higher-order eccentricity terms,
\bd
\sigma_1= -0.46 \frac{m_2}{M+m_1}, \hspace{1cm}
\sigma_2= 3.0 \frac{m_1}{M+m_2}.
\ed

We should consider as primary data only those numbers issued from the fit of the dual Keplerian model to the observations. 
They are given in Table \ref{tI}. 
Afterwards, they are completed with those given in Table \ref{tII}, which are derived from the primary parameters and some additional hypotheses as indicated in the footnote of the table.

\begin{table}[h]
\caption{Parameters determined from the observed radial velocities. \citep{May}}
\begin{center}\begin{tabular}{lll}
\hline
parameter			&	planet c	&	planet b \\
\hline
$\eliK$ (m/s)		&	61.5 $\pm$ 1.7	&	45.8 $\pm$ 1.0	\\
Sidereal Period (days) &	219.4 $\pm$ 0.2	&	435.1 $\pm$ 1.4	\\
Periastron time 	(SJD$\dag$)&	52284 $\pm$ 1	&	51758 $\pm$ 13	\\
Eccentricity		&	0.38  $\pm$ 0.01	&	0.18 $\pm$ 0.04	\\
$\omega$			&	124 $\pm$ 3		&	237 $\pm$ 13	\\
\hline
$V_r$ (km/s)		& \multicolumn{2}{c}{8.144 $\pm$ 0.04}	\\
\hline
\multicolumn{3}{l}{$\dag$ SJD (simplified Julian date) $ = $ JD $ - $ 2,400,000}\\
\end{tabular}\end{center}\label{tI}\end{table}

\begin{table}[h]
\caption {Parameters derived from those in Table \ref{tI}. Epoch= SJD 52,000}
\begin{center}
\begin{tabular}{lll}
\hline
parameter	&	planet c	&	planet b \\
\hline
$\sigma\ddag$	($10^{-3}$)&	-0.7	&	4.6	\\
T  (days) (osculating)	&	219.2	&	437.1	\\
Mean Longitude at the epoch (deg.)&	18  &   76 \\
$a$ (AU) (osculating)	&  0.746	&	1.18	\\
$f(M,m,i)$ ($10^{-9} M_\odot$)	&	4.186		&	4.113	\\
$m \sin i$ ($M_\mathrm {Jup}$)$\dag$   & 1.86	& 1.85 \\
$m \sin i$ ($10^{-3} M_{\star}$)$\dag$   & 1.54	& 1.53 \\
\hline
\multicolumn{3}{l}{$\ddag$ assuming $i \approx 90^\circ$}\\
\multicolumn{3}{l}{$\dag$ assuming for the star mass $M_\star=1.15 M_\odot$ }\\ 
\end{tabular}\end{center}
\label{tII}
\end{table}

One critical additional datum introduced to obtain Table \ref{tII} is the mass of the star. We have adopted the mass $M_{\star} = 1.15 \pm 0.09 M_\odot$ as determined by \citet{May} on the basis of the Geneva stellar evolutionary models. However, other values exist in the recent literature: $M_{\star} = 0.93 \pm 0.09 M_\odot$ \citep{All}, $1.05 M_\odot$ (Geneva planet search web page, July 31th, 2002) and $1.11 M_\odot$ \citep{Che}. 
Several parameters given in Table \ref{tII} are very sensitive to the adopted stellar mass. 
Their values are shown for two different masses in Table \ref{tIII} for the sake of giving an idea of the inaccuracies due to the insufficient knowledge of the stellar mass. 
These differences indicate that the errors in the considered parameters are much larger than those that we would obtain by applying usual rules to derive them from the errors given in Table \ref{tI}. 
It is worth mentioning that the osculating semi-major axes and periods, the mass functions and the ratio of two planetary masses have only a weak dependence on the chosen star mass.

\begin{table}[h!]
\caption {Parameters showing a dependence on the adopted stellar mass}
\begin{center}
\begin{tabular}{lllll}
\hline
&\multicolumn{2}{c}{$M_{\star}=1.15 M_\odot$ }
&\multicolumn{2}{c}{$M_{\star}=1.05 M_\odot$ }\\ 
\hline
parameter  &  planet c	&	planet b &	planet c	&	planet b \\
\hline
$a$ (AU)$\dag$	&  0.746	&	1.18&0.72 & 1.15\\
$m \sin i$ ($M_\mathrm {J}$)   & 1.86	& 1.85 & 1.75 & 1.74 \\
$m \sin i$ ($10^{-3} M_{\star}$)   & 1.54	& 1.53 & 1.59 & 1.58 \\
\hline
\multicolumn{5}{l}{$\dag \ $ Osculating}\\
\end{tabular}\end{center}
\label{tIII}
\end{table}

For completeness, we should mention that given the small velocity of the system, 
$\sim 8$ km/s, and the short time span of the observations, the apparent and true time scale are not significantly different and, thus, we do not need to consider the ``planetary'' aberration.

\section{On the 3-body (dynamical) fit}

\begin{figure}[t!]
\epsscale{.80}
\plotone{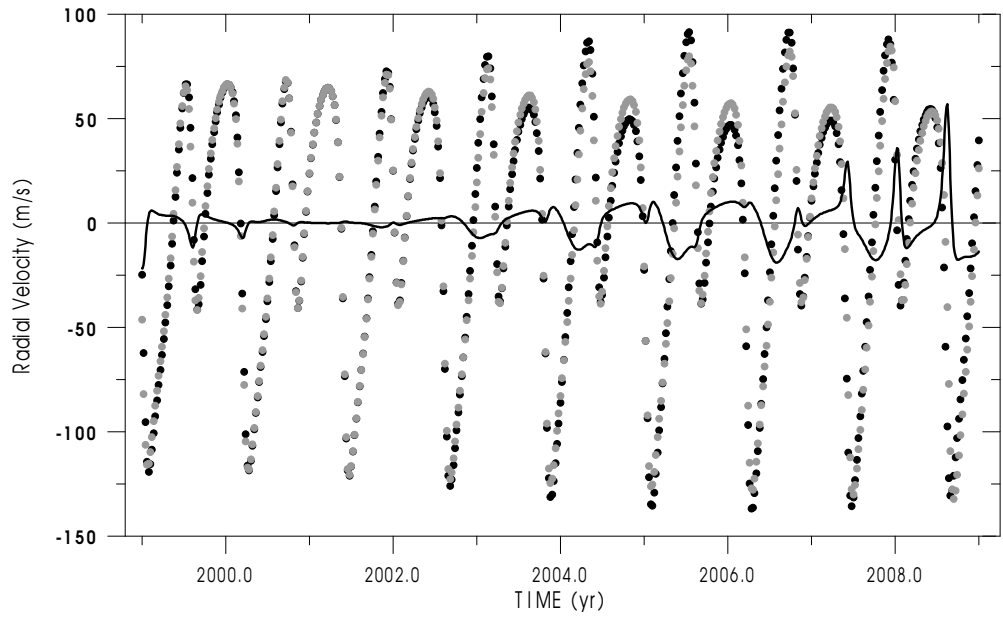}
\caption{Radial Velocity in kinematical (gray dots) and dynamical (black dots) models corresponding to \citet{May} data and coinciding at the date 2001.247. The solid line shows the difference kinematical$-$dynamical. The existing observations lay in the interval $1999.0-2003.4$.}
\label{vradcomp}
\end{figure}

It is often said that a 3-body (dynamical) fit is significantly better than the fitting of the dual Keplerian (kinematical) model to the observations. 
However, this is not yet the case for the planets HD 82943 c,b. 
This system is much wider than GJ 876 for which successful determinations with a 3-body model have been possible \citep{Lau}. 
Therefore, the mutual attraction of the planets is much smaller than in GJ 876 and the periods are much larger: the observational time span of HD 82943 corresponds to less than 4 periods of the outer planet. 
As a result, the radial velocity curves obtained with the dynamical approach or the kinematical one (which correspond to neglect the planet masses in the interaction term) are very similar. 
They are given in Fig. \ref{vradcomp} and show that a reasonable increase in the observation time span is necessary before getting results different from these models. 
It is worth mentioning that the good observational season for the star HD 82943 is the period from November to June and that observations done in the latest season were not included in our sample. 

Anyway, we have also run some fits using a coplanar 3-body model. 
The indetermination involving $M,m,i$ in the kinematical approach is not completely solved by the dynamical approach: the star mass and the orbital inclination continue entangled inside the classical gauge $M_{\star}\sin i$. 
Since the ratio of the two planetary masses is a robust quantity (not affected by the indetermination in the star mass), it is convenient to substitute one of the planetary masses by the mass ratio (Ferraz-Mello et al., in preparation). 

The introduction of the gauge and the remaining planetary mass, as independent unknown quantities, in a differential correction procedure has resulted in a correlation factor $\sim 0.99$ between the derivatives with respect to the two unknowns. 
This shows that, with the current observations, it is not yet possible an independent determination of the planetary masses, exactly as the radial velocity curves of fig. \ref{vradcomp} led us to expect.

Therefore, in the next sections, we will always use the kinematical model, which makes the calculations some orders of magnitude faster than the dynamical model and, thus, allows a thorough analysis to be more easily done.

\section{Good-fit solutions}\label{goodfit}

\begin{figure}[t!]
\epsscale{.80}
\plotone{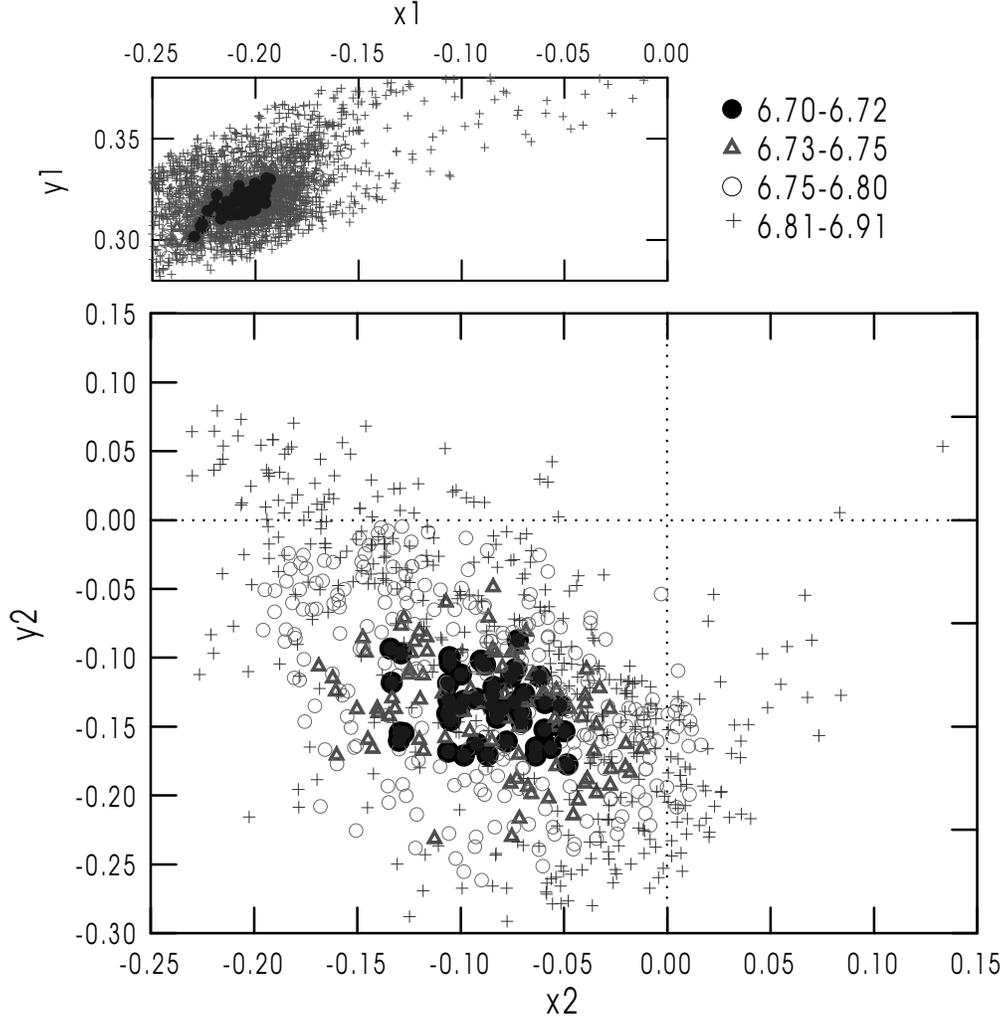}
\caption{Initial conditions generated by means of a biased Monte Carlo technique. The main panel shows the projections on the plane $x_2=e_2\cos\omega_2, y_2=e_2\sin\omega_2 $ of the initial conditions whose corresponding solutions have residuals with a r.m.s. less than 6.91 m/s.
The inset on top of the figure shows the projections of the same solutions on the plane $x_1=e_1\cos\omega_1, y_1=e_1\sin\omega_1 $. }
\label{monte}
\end{figure}

Several authors have insisted on the non-linear nature of the problem under study and have used or commented on the possible use of genetic algorithms to determine the minima of the function \citep{Lau,For,Goz}. 
In the case of the planets HD 82943 c,b, we followed, however, a different path. 
First, we have realized that almost every initial value leads, by a descent along the gradient, to the neighborhood of the minimum corresponding to the solution given by Mayor et al. 
Second, the population thus obtained, which should be used to start a genetic algorithm, revealed itself good enough to understand the problem. 
The final population, with 26,000 solutions, was obtained by means of a random (Monte Carlo) sampling in a wide domain of the initial conditions, biased to keep only initial conditions leading to residuals with an r.m.s. value not larger than 10.0 m/s. 
The initial conditions corresponding to r.m.s values smaller than 6.9 m/s are shown in fig. \ref{monte}.
This figure presents each solution in the plane $x_2=e_2\cos\omega_2, y_2=e_2\sin\omega_2 \,$[{\footnote {The subscripts 1 and 2 refers to the inner and outer planet, respectively.}}], but it is important to keep in mind that each set of initial conditions is complete. Each point in fig. \ref{monte} is, thus, the projection on the $x_2,y_2$ plane of a point in an 11-D space. The variables chosen for this projection, $x_2,y_2$, are the variables ill constrained by the observations and whose values are critical for the stability of the motion. 
We may compare the distribution of these points with that of the projections of the initial conditions on the plane $x_1=e_1\cos\omega_1, y_1=e_1\sin\omega_1 $ shown in the inset on top of fig.\ref{monte}, which are much more concentrated. 

The analysis of the results shown in figure \ref{monte} involves several complementary aspects. 

First, the ($O-C$) corresponding to the solutions shown in fig.\ref{monte} do not differ significantly from the ($O-C$) of the best-fit solution.
A straightforward calculation \citep[see][chap. 14]{Rec} allows us to estimate that they belong to the confidence region corresponding to the level 38.6 percent -- the $\sigma/2$ level. (The confidence region corresponding to the standard $1\sigma$ level would include all solutions with r.m.s less than 7.1.) 
This estimate is founded on several hypotheses: 
a) The distribution of the errors is normal; 
b) The best-fit solution corresponds to the expected value of the quantity $\sum(O-C)^2$ (a $\chi^2$ variable); 
c) $\sum(O-C)^2$ is always the sum of two orthogonal $\chi^2$ variables. 
These hypotheses are very stringent and certainly not rigorously fulfilled. 

At this point, we should say that it is culturally difficult to completely abandon the heuristics of the least squares. 
However, we have to keep in mind that our ultimate goal is not to look for the minimum of a mathematically defined function, but for initial conditions that reproduce equally well the observed values. 

\begin{figure}[h]
\epsscale{.80}
\plotone{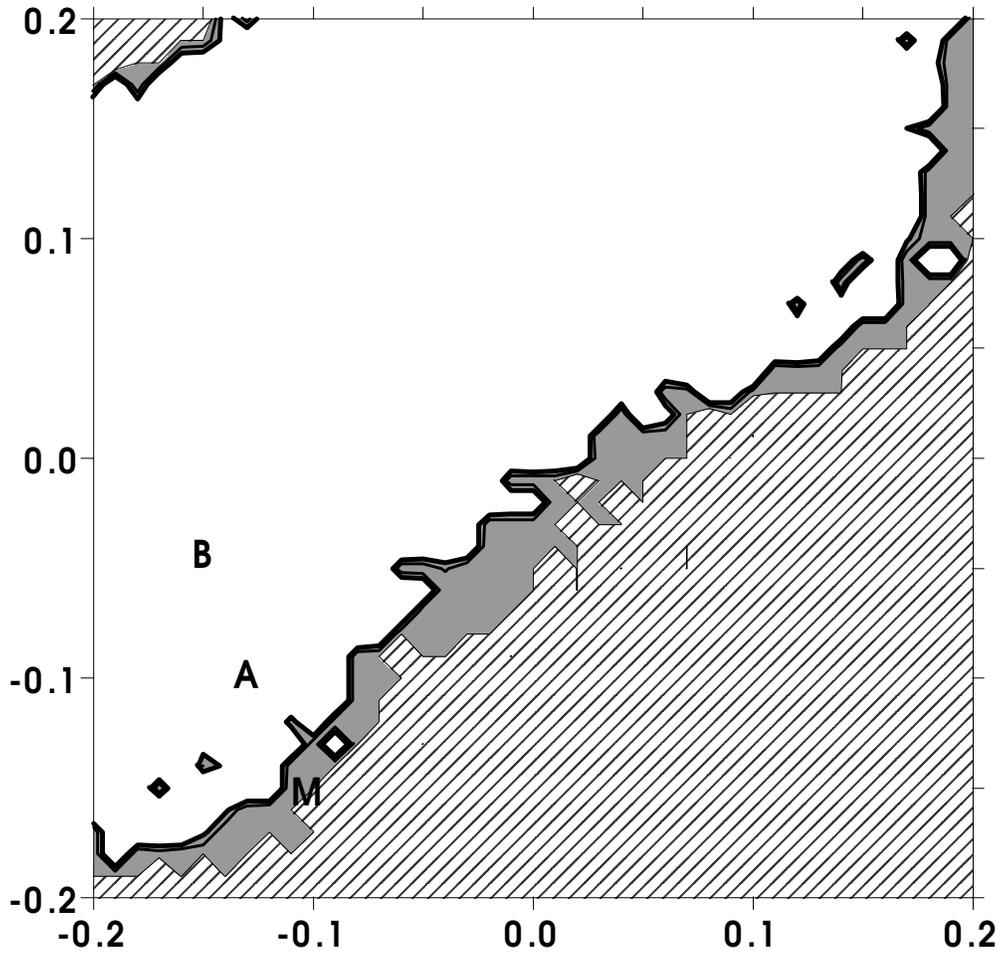}
\caption{Collision-regularity diagram in the plane $x_2,y_2$, for the case $\sin i=1$. The hachured region correspond to initial conditions leading to collision in less than 260,000 yrs. The gray region corresponds to solutions surviving that time span but showing strong chaos. In the white region, solutions appear as regular. Letters indicate the solutions listed in table tIV.}
\label{caos}
\end{figure}

Second, most of the good-fit solutions thus obtained appear grouped in the left lower quarter of the plane. We may compare the position of these points and those showing the projections on the plane $x_1=e_1\cos\omega_1, y_1=e_1\sin\omega_1 $ (shown in the inset of fig. \ref{monte}). 
It is well known that anti-aligned periapses and high eccentricities in a 2/1-resonant system are ingredients generally leading to instability. 
Fig.\ref{caos} shows the fate of solutions whose initial conditions are those of solution B (see Table \ref{tIV}), except for the variables $x_2=e_2\cos\omega_2, y_2=e_2\sin\omega_2 $, which are let to vary in the domain shown in figure\ref{monte}. The hachured regions correspond to orbits bound to collision in less than 260,000 years. The gray strip correspond to orbits that survived the simulation time span but that are strongly chaotic and thus bound to a catastrophe in a not much larger time span.
(The diagnostic of chaos was done with the spectral technique discussed by \citet{Mag}.) 
A great proportion of the solutions shown in Fig. \ref{monte}, as well as the solution given by Mayor et al., lie inside the collision region or in the adjacent strip of chaotic solutions. Solutions in the white area behave as regular in the considered time span (260,000 years) and we may expect then to survive for long times.

\section{Some selected solutions}

We select for analysis two of the solutions shown in fig.\ref{monte}. The first of them (A) is a stable solution taken amongst the solutions with the lowest r.m.s.; it may last for many $10^8$ yrs if the system is edge-on. 
However, if the mean plane of motion has an inclination 45 degrees, the planet masses are $\sqrt{2}$ larger and the system disrupts in less than $10^8$ yrs.  
The same occurs if the orbits are nearly orthogonal to the sky plane, but have a mutual inclination of 5 degrees. The second selected solution (B) has a slightly higher r.m.s., but is very robust and remain regular even if the system has a 30-degree inclination over the sky plane, that is if the planet masses are 2 times larger than the minimal masses. These solutions are given in Table \ref{tIV} where elements are grouped by planet to allow us a better visualization of the differences in each element. For the same reason some less significant digits are kept in the given solutions. 

\begin{table*}[t]
\caption{Osculating elements and masses in some chosen solutions compared to those given by Mayor et al.
Epoch JD 2,452,000}
\begin{center}\begin{tabular}{lllllll|l}
\hline
Solution &  a    &  P     &   e    &   $\omega$   &   $\lambda_{\mathrm epoch}$   &   $m\sin i$ & r.m.s. \\
	&    (AU)	&	(days)	&	&	(deg.)	& (deg.)	& ($M_\mathrm {Jup}$) &	(m/s)\\
\hline	
\multicolumn{6}{c}{Planet c (inner)} && \\
\hline 
A&	0.746&	219.44&	0.386&	125&	18.2&	1.78&	6.72\\
B&	0.746&	219.54&	0.396&	123&	17.6&	1.71&	6.76\\
Mayor et al.$\dag $& 0.75& 219.4&	0.38&	124&	18  &	1.85\\
\hline	
\multicolumn{6}{c}{Planet b (outer)}&& $m_2/m_1$\\
 \hline 
A&	1.181&	436.90&	0.161&	217&	76.2&	1.84&	1.04\\
B&	1.180&	436.17&	0.153&	196&	76.1&	1.82&	1.06\\
Mayor et al.$\dag $& 1.18& 435.1&	0.18&	237&	76  &	1.84&	\\
\hline
\multicolumn{7}{l}{Adopted star mass: 1.15 $M_\odot$}\\
\multicolumn{7}{l}{$\dag \ $ Not osculating. Indicated with M in figs. \ref{caos} and \ref{fig8}.}\\
\end{tabular}\end{center}\label{tIV}\end{table*}

\begin{figure}[h]
\epsscale{.80}
\plotone{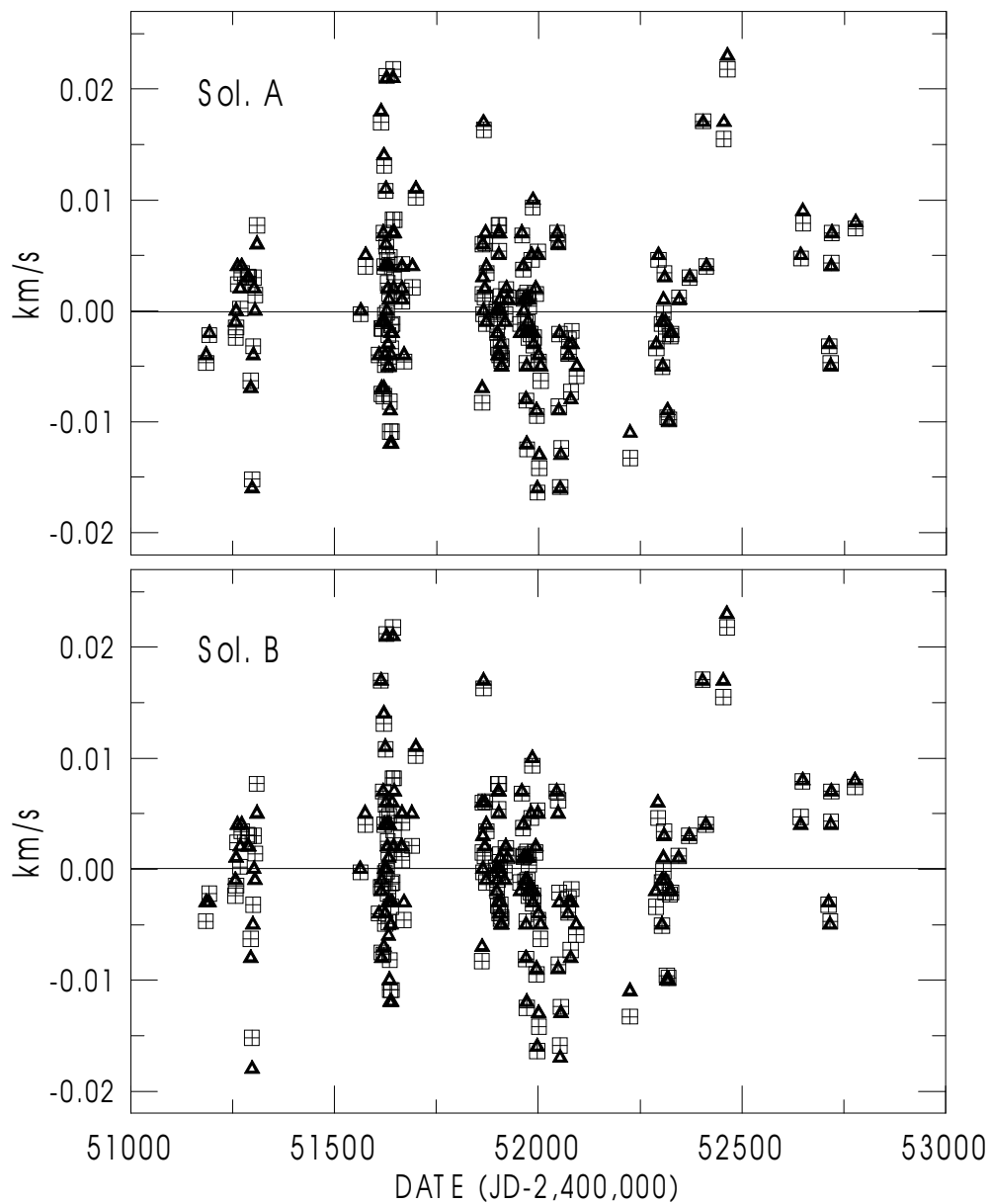}
\caption{$O-C$ of the solutions given in Table \ref{tIV}. The crossed squares are the $O-C$ given by Mayor et al. The triangles are the $O-C$ of the solutions A and B.}
\label{OmC}
\end{figure}

Notwithstanding some differences in the elements and masses, the $O-C$ residuals change only slightly. They are shown in fig.\ref{OmC} where, for the sake of comparison, the $O-C$ taken from Mayor et al. for the same dates are shown. 
Only a thorough inspection of the plots can show differences among them.
These figures show that solutions A and B are good candidates to be the right orbits and that it is impossible to choose one of them to be labeled as {\it the} orbits of the planets HD 82943 c,b. 
The same is true for many other solutions shown in fig.\ref{monte}.

\begin{figure}[h]
\epsscale{.80}
\plotone{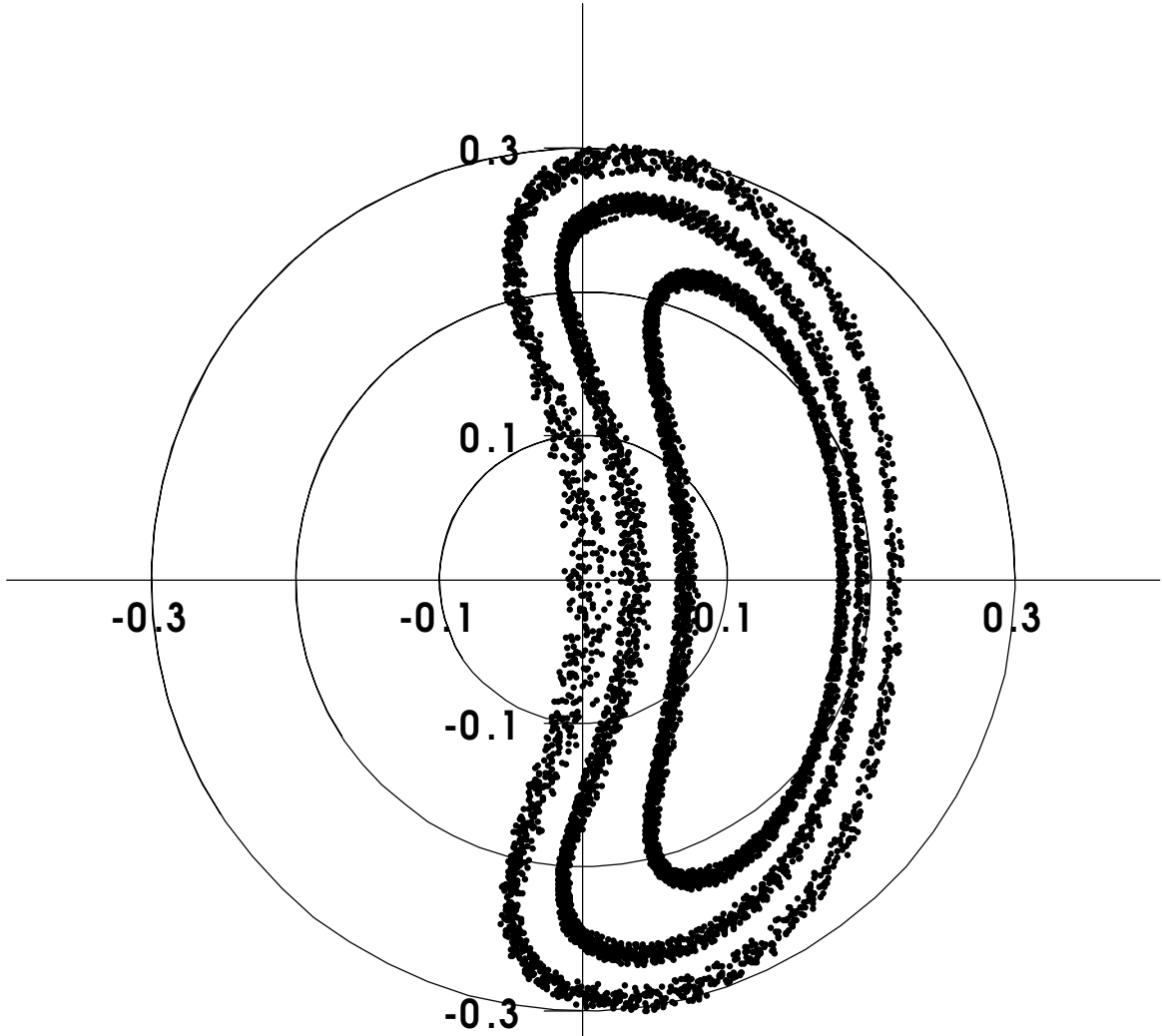}
\caption{Joint variation of the eccentricity of the outer planet and the difference of the periastron longitudes of the two planets in three solutions: the less unstable among the discarded least-squares solution (with r.m.s.= 6.70 m/s) and the two selected solutions A and B. Polar coordinates: radius vector: $e_2$, polar angle: $\Delta\omega=\omega_b-\omega_c$. From inside to outside: solution B, solution A and the discarded least-squares solution.} 
\label{edpi}
\end{figure}

We may proceed and make a study of the selected solutions. 
The most conspicuous results concern the ill determined eccentricity and periastron longitude of the outer planet and the critical angles of the 2/1 resonance. 
The joint variation of the eccentricity of the outer planet and the difference of the periastron longitudes of the two planets are shown in fig. \ref{edpi} (the two innermost curves). 
For comparison we add the curve corresponding to the less unstable of the least-squares solutions (r.m.s = 6.70 m/s).
The important result we get from it is that these solutions are similar and only differ in the amplitude of oscillation of the angle $\Delta\omega=\omega_b-\omega_c$. 
The virtual center of the loops is a (0,0)-apsidal corotation (aligned periapses) with eccentricities $e_{10}\sim 0.13, e_{20}\sim 0.4$, a pair of values consistent with two planets of very similar masses \citep{Be1,Be2}. 
The loops corresponding to solutions A and B do not enclose the origin. 
So, in these cases, the two periapses are oscillating one with respect to another, without circulation. 

\begin{figure}[h]
\epsscale{.80}
\plotone{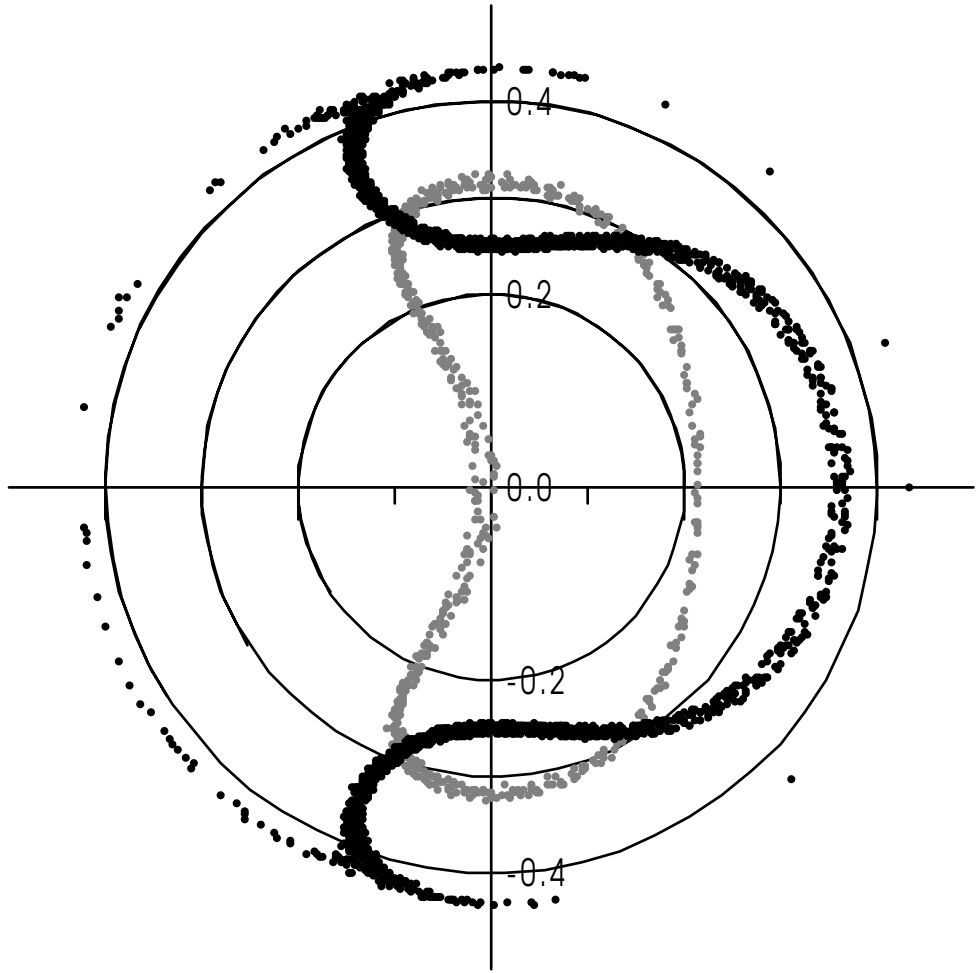}
\caption{Joint variation of the eccentricities and the difference of the periastron longitudes of the two planets in the solution given by \citet{May}. Polar coordinates: radius vector: $e_1$ (black) and $e_2$ (grey); polar angle: $\Delta\omega=\omega_b-\omega_c$.}
\label{leak}
\end{figure}

The curve corresponding to the outer planet in the solution given by Mayor et al. (inner curve in fig. \ref{leak}) is very similar to the outer curve in fig.\ref{edpi} (corresponding to a least-squares solution).
In both cases, the curves enclose the origin. 
It is clearly seen that the left side of the loop crosses the x-axis at the left of the origin.
However, the fact that periapses are aligned at that moment (the curve passes at the right of the origin) or anti-aligned (the curve passes at the left of the origin) has no great signification because the corresponding eccentricity is very close to zero (the outer orbit is almost circular). 
In such case, there are no dynamical differences arising from the fact that the periapsis is oscillating or circulating. 
The difference is purely geometric (kinematical) without any topological (dynamical) meaning.
It is worth mentioning also that when the loop of the outer planet encloses the origin, but passing very close to $e_2=0$, the loop of the inner planet "leaks" (see fig.\ref{leak}). 
It circulates in high eccentricity but, again, this is dynamically meaningless because this happens when $e_2\sim 0$ and the periapsis of the outer planet is ill defined; the orbit of the outer planet is almost a circle and the mathematical location of its periapsis is immaterial. 
This is certainly not the reason for which the least-squares solution is an unstable orbit while our solutions are not. 

\begin{table}[h!]
\caption {Libration Half-Amplitudes}
\begin{center}
\begin{tabular}{lcccl}
\hline
Solution	&	in $\theta_1(^\circ)$	&	in $\theta_2(^\circ)$ & Period (yrs)	\\
\hline
A	&	43	&	57	&	$\sim$620 	\\
B	&	72	&	89	&     $\sim$630	\\
\hline
\hline
\end{tabular}\end{center}
\label{tV}
\end{table}

We may draw figures similar to fig. \ref{edpi} considering the 2/1-resonance critical angles $\theta_1=2\lambda_2-\lambda_1-\omega_1$,$\ \theta_2=2\lambda_2-\lambda_1-\omega_2\,$[\footnote{We have written $\omega_i$ instead of $\varpi_i$ as usual in the definition of these variables, since in coplanar models the nodal line is arbitrary and the two angles may be considered equal.}], instead of $\Delta\omega$. In the 2 solutions, these angles librate about $\theta_i=0$ with the half-amplitudes shown in Table \ref{tV}.
Again, when the eccentricity of the outer planet becomes very close to zero, the angles $\theta_1$ and $\theta_2$ lose dynamical meaning. In fact, in all calculations done in this paper the eccentricities and arguments of periapses were substituted by the variables $x_i,y_i$ introduced in section \ref{goodfit}.

\section{Effects due to inclination}

The primary data arising from the kinematical orbit determination do not include the plane inclinations (nor the planet masses). 
However, dynamics simulations depend on the unknown planet masses and the mutual inclination of the planes.
The values of these quantities need to be assumed.
For instance, the simulations used to construct fig. \ref{caos}, were done assuming that the orbits are coplanar and $i=90^\circ$.
In what follows, we will consider separately the effects of $\sin i \ne 1$ and the non-coplanarity of the orbits of the two planets.

\subsection{Coplanar orbits not edge on}

\begin{figure}
\epsscale{.90}
\plotone{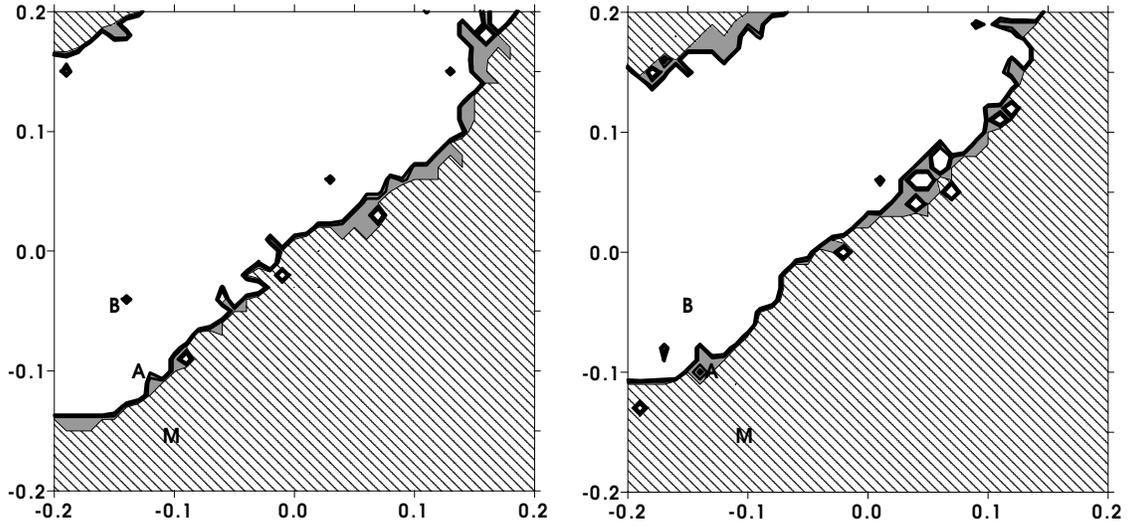}
\caption{Same as fig. 4 for $1/\sin i=\sqrt{2}$ and $1/\sin i=2$.}
\label{fig8}
\end{figure}

If the planet orbits are not edge on, their masses are larger than those assumed to draw fig. \ref{caos}. 
It is then important to know how the given result change when $\sin i\ne 1$.
Fig. \ref{fig8} shows two plots constructed in similar way, assuming that the planets are coplanar, but inclined with respect to the tangent plane to the celestial sphere. The adopted inclinations in these figures are $i=45^\circ$ and $i=30^\circ$, respectively. Therefore, the masses of the planets are larger than in the previous case, by factors $\sqrt{2}$ and 2, respectively.

The resonance zones are, now, narrower and conditions for stability are more stringent. 
In the first case ($i=45^\circ $), one of the solutions discussed previously (solution A) is close to the border of  the regularity zone and becomes unstable in relatively short times ($\sim 10^8$ yrs). In the second case ($i=30^\circ $), this solution is engulfed in the collision region. 
Solution B remains in the regularity zone in both cases. 

A simulation of solutions in the same situation as solution C, with $i=30^\circ $ ($1/\sin i=2$) showed the same libration pattern and amplitude for longer than $10^9$ years.

\subsection{Mutually inclined orbits}

We have done several simulations with the planets in mutually inclined orbits using the RA15 code of \citet{RA}.
Mutual inclination contributes to increase the average distance between the two planets affecting stability, and we should know how it changes the stability of the best-fit solution before looking for alternative fits.
The question is of how inclined the orbits can be. 
All cosmogonics discussions are in favor of almost coplanar orbits. 
Anyway, we have explored inclined orbits up to $10^\circ$ with all possible choices of the longitude of nodes.
In all experiments, the reference plane was Laplace's invariant plane, which is orthogonal to the total angular momentum of the system:
\be
{\cal L} = m_1 n_1 a_1^2 \sqrt{1-e_1^2} \cos i_1 + m_2 n_2 a_2^2 \sqrt{1-e_2^2} \cos i_2. \label{momang} \ee
The order ${\cal O}(m_i^2)$ correction of these equations due to the fact that we are using osculating Keplerian elements was neglected \citep[For a rigourous formulation, see ][]{M05}. 
The inclinations and nodes are thus chosen such that the component of the angular momentum in the reference plane is zero: The two nodes are $180^\circ$ apart one from the other, and
\be
m_1 n_1 a_1^2 \sqrt{1-e_1^2} \sin i_1 = m_2 n_2 a_2^2 \sqrt{1-e_2^2} \sin i_2.
\ee
In our simulations, the choice of the individual nodes and periapses was done is such a way that, for each planet, the resulting distance $\omega$ of the periastron to the sky tangent plane is the value given by the orbit determination. 
Some results posted in Internet, obtained with arbitrary node longitudes simply added to the distances $\omega_i$, do not correspond to the actual planets since the given $\omega_i$ are arguments of periastron only when the reference plane is the sky tangent plane. 
When the arbitrary addition of a longitude of node is done for another reference plane, the resulting orbits no longer correspond to the observed ones.

The results of simulations done with the orbital elements and masses given in tables \ref{tI} and  \ref{tII} and mutually inclined orbits remained stable for 1--2 million years, but not longer. Only one anomalous case exceeded these values and remained stable for the whole duration of the simulation (10 Myr). However, it was not possible to reproduce it with a faster computer (the same initial conditions and codes leading, then, to a catastrophic event in less than 1.5 Myr). Besides, in a test of robustness with the same computer, a small change of some $10^{-3}$ degrees in the position of the reference plane was enough to give rise to instability, again, in less than 1.5 Myr. 
That long lasting anomalous solution, frankly chaotic, seems rather due to sticking to an artificial stability island. The exploration of a grid of initial conditions with mutual inclinations $i<30^\circ$ did not show any real island of stability. 

\begin{figure}
\epsscale{.70}
\plotone{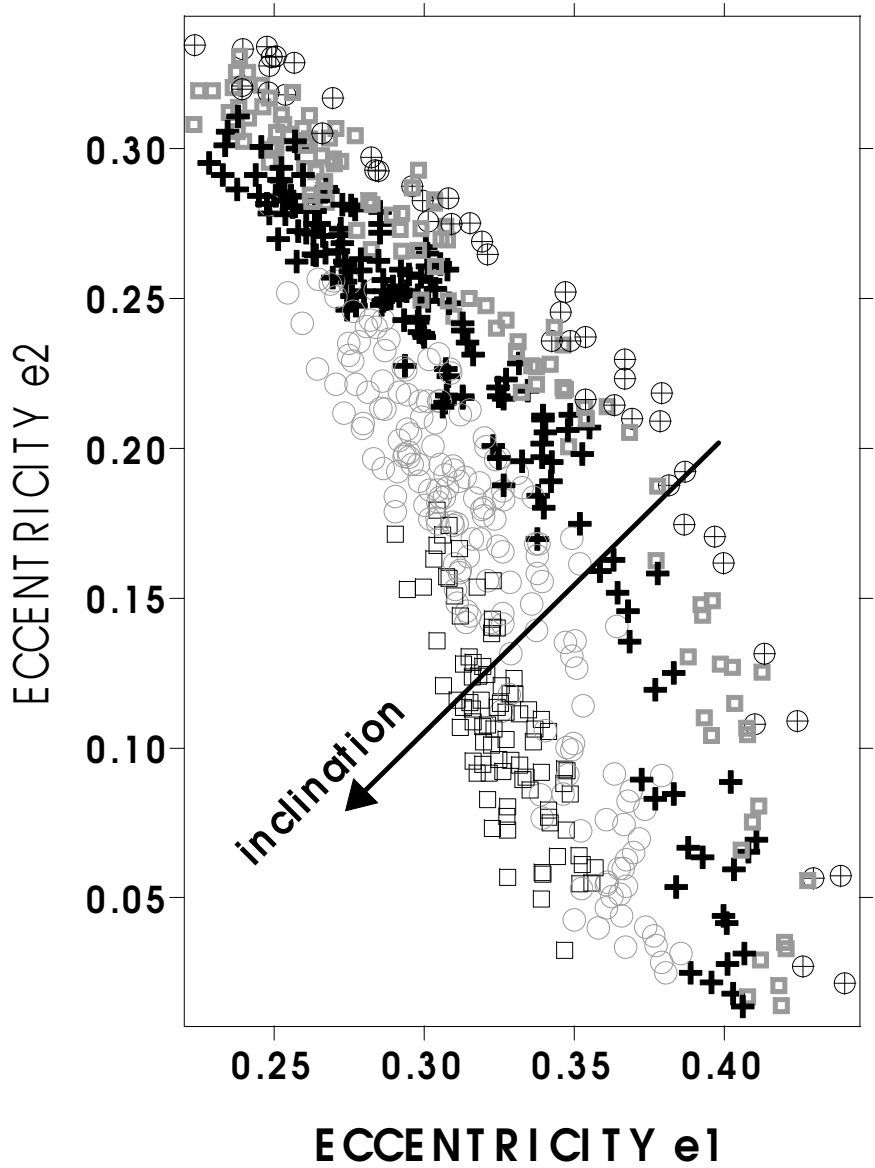}
\caption{$e$-$e$-$i$ coupling during the stable phase of one simulation with initial mutual inclination $10^\circ$. The symbols (alternately black and gray) correspond to the values of the mutual inclination. From right to left: mutual inclination ranges $6-10^\circ$, $\,10-14^\circ$, $\,14-18^\circ$, $\,18-22^\circ$ $\,22-26^\circ$.}
\label{eei}
\end{figure}

One last consideration concerning the stability of mutually inclined orbits follows from the analysis of eqn. (\ref{momang}). First, let us introduce
\be
{\cal L}_{\|} = m_1 n_1 a_1^2 \sqrt{1-e_1^2} + m_2 n_2 a_2^2 \sqrt{1-e_2^2}
\ee
and write eqn. (\ref{momang}) as
\be
{\cal L}_{\|} = {\cal L} + \Delta {\cal L}
\ee
where
\bd
\Delta {\cal L} = 2m_1 n_1 a_1^2 \sqrt{1-e_1^2} \sin^2{\frac{i_1}{2}} + 2m_2 n_2 a_2^2 \sqrt{1-e_2^2} \sin^2{\frac{i_2}{2}} \ge 0.
\ed 
It is known that the integral ${\cal L}_{\|}$ controls the $e$-$e$ coupling in a coplanar motion limiting the variations of $e_1$ and $e_2$, which may remain below some limits (the limit for one eccentricity is reached when the other is equal to zero and cannot be exceeded as far as the semi-major axes are kept almost constant). If the coplanar motion corresponding to a set of elements is unstable because of eccentricities able to reach large values, the existence of a mutual inclination acts worsening this condition. Indeed the 3-D integral ${\cal L}$ controls a similar ${\cal L}_{\|}$-$i$ (or $e$-$e$-$i$) coupling (see fig \ref{eei}). 
When the inclination grows, both $\Delta {\cal L}$ and ${\cal L}_{\|}$ grow and improve the stability of the system (the maximum allowed eccentricities become smaller). But the converse is also true. When the inclination decreases, both $\Delta {\cal L}$ and ${\cal L}_{\|}$ decrease allowing larger eccentricities to be reached, which may lead the system to become unstable. 
This phenomenon was observed in the simulations with an initial non-zero mutual inclination. 
The system remains stable as far as the mutual inclination remains high, but no bounds exist and, sooner or later, it becomes small and the system is bound to a catastrophic event. 

\section{Conclusions}

We have shown that several sets of planetary elements and masses exist satisfying the observational data of HD 82943 almost as well as the least-squares solutions. The goodness of fit of these solutions may be estimated by using some $\chi^2$ rules, but better than that -- since we cannot guarantee that the conditions for the use of $\chi^2$ statistics are fulfilled -- is the comparison of the $O-C$ plots of these solutions. The proposed solutions are regular and correspond to stable planetary systems, at least as far as we may assume that the orbits are coplanar and being seen edge-on, at variance with the short lived system that corresponds to the published orbits of the planets HD 82943 c,b. If the planets have different inclinations and the masses are significantly larger than assumed, solution A becomes unstable and bound to a catastrophic close approach of the two planets. 

The analysis of the available observational data shows that this planetary system is trapped in a 2/1 mean-motion resonance and the apsidal lines of the two orbits oscillate one around another, keeping the system in the neighborhood of a (0-0)-apsidal corotation. 
These orbits suggest a capture into resonance not followed by the complete damping of the libration amplitudes. We may compare this system to the GJ 876 planetary system, whose libration amplitudes are close to zero. Either the damping in HD 82943 was not efficient enough or it did stop long before a steady state was reached. Since the size of this system is larger than that of GJ 876 (the orbits of the outer planet in the two cases are, respectively, 1.18 and 0.21 AU), we may guess that the larger size affected the relative torques acting on the planets and did not allow the system to be damped to a final stationary solution.

\begin{figure}
\epsscale{.80}
\plotone{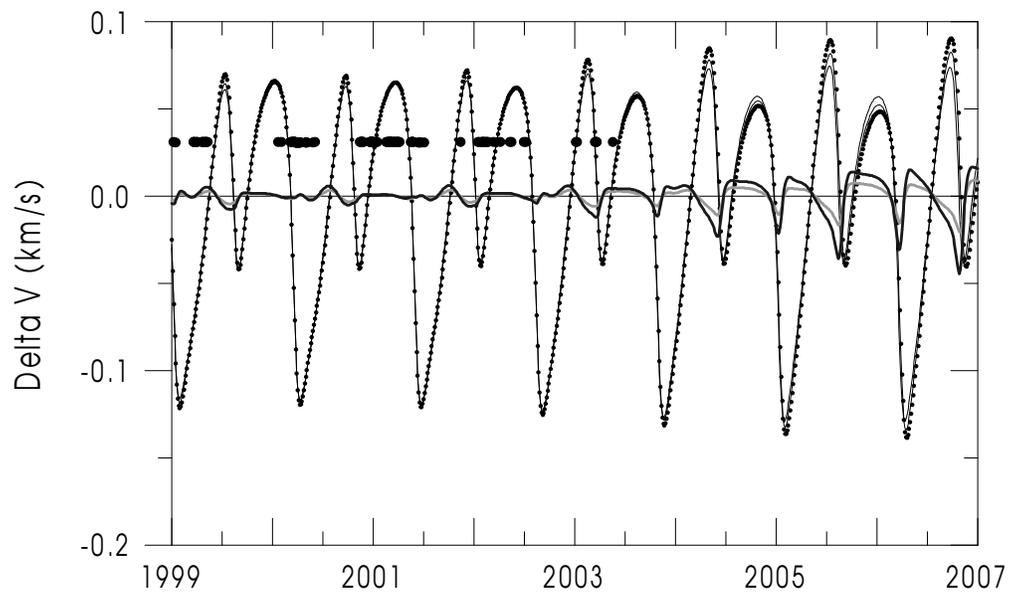}
\caption{Radial velocity curves corresponding to the solutions discussed in this paper. The thick solid lines show the differences of solutions A(gray) and B(black) to the solution given by Mayor et al (dotted curve). The wide dots show the dates corresponding to the available data.}
\label{fig10}
\end{figure}

One important question concerns the possibility that new observations allow the real orbit to be known. Fig. \ref{fig10} shows the curves of radial velocities and their differences to the curve corresponding to the solution published by Mayor et al. The differences show that new observations may help getting best-fit solutions corresponding to non-colliding orbits. However, some years will elapse before the observations allow us to distinguish between the proposed solutions. Just for completeness of the information given, we mention that the differences were obtained forcing the same period in all orbits to be sure that they correspond to real features of the radial velocity curves and not to phase shifts. 

Another point to be stressed here is the separation of the primary elements obtained in the orbit determination from those including additional parameters as the mass of the central star and the inclination (Section 2). In particular, we mention that the observed periods are not osculating and that they should not be used as initial conditions in simulations. The recipes to obtain the osculating period and semi-major axes are however simple (as indicated in the appendix). 

\acknowledgements
{This investigation benefited from grants from Brazilian Research Council -- CNPq -- Procs. 300575/90-4(SFM), 300953/01-1(CB) and 303196/02-5(TAM), the Sao Paulo State Research Foundation -- FAPESP -- Procs. 98/15431-5(SFM) and 00/07074-0(CB) and from Conicet. 
The authors gratefully acknowledge the support of the Computation Center of the University of S\~ao Paulo (LCCA-USP) for the use of their facilities.}

\appendix
\section{Appendix. How mutual perturbations affect the periods}
This is a classical subject in Celestial Mechanics, known since Laplace (at least), which has not been considered up to now in papers on extra-solar planetary systems. 
Besides periodic variations, mutual perturbations affect the mean value of osculating elements. 
Two of these variations are particularly important: the mean perturbations in semi-major axis and the mean longitude at the epoch. 
The latter is a secular drift of that angle which directly affects the period of the motion. 
The classical first-order formulas giving these effects are the following
\citep[][chap.20]{Tis}\citep[see also][sec.5.4]{sfm}: 
\be
\bar{a}_k = <r_k> = a_k ( 1-\frac{1}{2} \sigma_k), \hspace{1cm}
\tilde{T}_k = T_k(1-\sigma_k)
\ee
where, as already mentioned, $a_k,T_k$ are the osculating semi-major axis and period, $\tilde{T}_k$ is the sidereal period and $\bar{a}_k$ is the mean distance from the star to the planet. 
Assuming that $a_1<a_2$ and neglecting higher powers in the eccentricity, we have 
\bd
\sigma_1=\frac{m_2 }{M+m_1}\alpha^2\,\frac{db^0_{1/2}}{d\alpha},
\ed
\be
\sigma_2=- \frac{m_1}{M+m_2}\,(\alpha\frac{db^0_{1/2}}{d\alpha} + b^0_{1/2});
\ee
where $\alpha=\frac{a_1}{a_2}$ and $b^0_{1/2}(\alpha)$ is the lowest order Laplace coefficient. 
For $\alpha \approx 0.636$ (as in the 2:1 resonance), we have $b^0_{1/2} \approx 2.268$, $db^0_{1/2}/d\alpha \approx 1.132$ and
\bd
\sigma_1= -0.46 \frac{m_2}{M+m_1}, \hspace{1cm}
\sigma_2= 3.0 \frac{m_1}{M+m_2}.
\ed

\end{document}